# Comment on Shadow and Non-Shadow Extensions of the Standard Model

Hitoshi Nishino[1)] and Subhash Rajpoot[2)]

*Department of Physics & Astronomy*
*California State University*
*1250 Bellflower Boulevard*
*Long Beach, CA 90840*

**Abstract**

The models in the two papers `hep-ph/0608068` and `hep-ph/0701254` by Chang *et al.* with the so-called shadow gauge and scalar fields are nothing but convenient tailored versions of our model in `hep-th/0403039`. The same remarks applies to the work in `hep-th/0612165` by Meissner and Nicolai.

---

[1)] E-Mail: hnishino@csulb.edu
[2)] E-Mail: rajpoot@csulb.edu

One of the most important issues in elementary particle physics is to seek explanation of the origin of masses of quarks, leptons and other fields. To this end, we extended in [1] the standard model in such a way so as to retain the standard model construct as well as address the issue of the problem of mass. The work dwells on earlier work by one of us [2].

We begin by recapitulating the most salient features of the our extension [1] of the standard model:

(i) It has local scale invariance that was considered by Weyl many years ago for different reasons [3].

(ii) Since mass and gravitational interactions are ultimately linked *via* Newton's law, we included gravitational interactions in such a way that Einstein-Hilbert action follows after spontaneously breaking of scale invariance [1].

(iii) The Higgs sector is extended to include a real scalar singlet [1].

(iv) The fermion spectrum is extended to include right-handed neutrinos [1].

(v) The mass spectrum of the quarks and leptons and other fields is generated by explicit spontaneous breaking scale invariance [1].

(vi) In our work the gauge field associated with local gauge invariance is called the Weylon, named in honor of the proponent of the gauge principle. It becomes massive after scale symmetry breaking [1].

(vii) The Weylon has either no or feeble couplings to the quarks and leptons [1].

(viii) Neutrinos acquire masses *via* the see-saw mechanism. The see-saw scale is, of course, provided by the scale associated with the primary descent [1].

We now comment on the works by Chang-Ng-Wu [4][5], and by Messner-Nicolai [6] in the light of the aforementioned features:

**(1) Chang-Ng-Wu's First Paper [4]**

In C-N-W's first paper [4], the authors take our model [1], excise scale invariance (item **(i)** above), excise gravitational interactions (item **(ii)**), excise right-handed neutrinos (item **(iv)**), and convert our real scalar singlet to a complex scalar singlet to present their work. The gauge field associated with local invariance which we call the Weylon is their so-called shadow extra Z-boson. Notice that their $U(1)_S$ is introduced in an *ad hoc* manner. In our work, the Weylon does not directly couple to the quarks and leptons which is the consequence of local scale invariance combined with general relativity. C-N-W assume this important property for their so-called extra shadow Z-boson associated with their $U(1)_S$ symmetry [4]. Furthermore, it is a well-known fact from standard model Z-boson physics that the coupling between any extra Z-boson and the Z-boson of the standard model is tiny, making the extra Z-bosons much heavier than the standard Z-boson, if no fine-tuning of the relevant couplings is allowed. In our model, since we were not presenting detailed phenomenological analysis, we set the mixing between the extra Z-boson and the Z-boson of the standard model to be zero, as the first go-round [1] in our analysis. C-N-W introduce such a coupling through the parameter $\epsilon$, and after an extensive phenomenological analysis, find it to be tiny in the end! The rest, such as conventional Higgs - leftover scalar singlet mixing phenomenology, are just details.

### (2) Chang-Ng-Wu's Second Paper [5]

C-N-W's second paper [5] is again our model [1] suitably tailored for specific purposes. This time, however, they *do* talk about local scale invariance [1]. Their $U(1)_S$ symmetry now becomes 'scale invariance' symmetry! Our scale invariance and the associated gauge field we call the Weylon are exactly the same as their shadow Z-boson and their $U(1)_S$ symmetry [4]. Again they excise the general relativity part of our model (item **(ii)**), convert our singlet scalar (item **(iii)**) [1] to a complex field, and excise right-handed neutrinos (item **(iv)**). Because of scale invariance, their scalar potential (eq. (19) in [5]) is nothing but the complexified version of our scalar potential (eq. (14) in [1]) with our scalar field replaced by a complex scalar. In our work [1], we break the scale symmetry explicitly. The authors in [4] break scale symmetry by radiative corrections.

In our work [1], the mass of the Weylon is given by $M_S = \sqrt{\frac{3f^2}{4\pi G_N}} \approx 0.5 \times f M_{Pl}$ (eq. (19) in [1]), where $f$ is the Weylon gauge coupling, and $M_{Pl}$ is the Planck mass. As a first go-round in our analysis, we took $f$ to be of $\mathcal{O}(1)$, so as not to run into problems with fine-tuning. However, the mass of the Weylon is arbitrary, since $f$ can be fine-tuned to be any value allowed by constraints coming from Z-boson physics.

It is unfortunate that the authors have neglected to cite our paper [1], and other references in the literature. It seems that the two separate works [4][5] have been generated out of the model in [1] as a premeditated effort.

### (3) Meissner-Nicolai's Paper [6]

First, we note that even though the title of their paper contains 'conformal symmetry', that symmetry is nothing but our scale invariance symmetry. The authors Meissner-Nicolai take our model, excise local scale invariance (item **(i)**) and the general relativity part (item (ii)). Their work is our model [1] with local invariance replaced by global scale invariance minus the relativity part (item (ii)). As we have stated before, we break scale invariance explicitly in our model, while M-N break it by radiative corrections. Notice that the scalar potential (eq. (1) in [6]) is nothing but our potential (eq. (14) in [1]). We present the Yukawa couplings in our paper [1] in a condensed form, while M-N write all the terms explicitly [6]. In any case, their basic lagrangian (1) in [6] is nothing but our lagrangian (15) in [1].

In conclusion, we revived [1] Weyl's scale invariance [3] so as to address the issue of mass in the standard model. Unfortunately for Weyl, time was not ripe, since the only interactions at his disposal were gravity and electromagnetism [3]. With the advent of the standard model and its unsurpassed successes, we considered it ripe to reconsider Weyl's proposal, as we have done in our work [1]. Weyl sought the unification of gravity and electromagnetism [3]. We seek to bring together all the interactions as well address the issue of particle masses [1]. The mere idea of extending the standard model with the addition of scalar singlets and gauge singlets is hardly new. But the idea of using a scalar singlet together with the Higgs doublet to break scale invariance and generate particle masses as we have done in our model [1] is original, and so is the introduction of a gauge field with feeble interactions with ordinary matter. It is in this spirit we incorporated Weyl's scale invariance [3] in the standard model [1]. The discovery of such a particle which we have dubbed the Weylon [1] at a future high energy facility will be yet another unquestionable testament to Weyl's great imagination [3]. Our original work [1] and its extension to scale invariant $SU(5)$ GUT is published in [7].

## References

[1] H. Nishino and S. Rajpoot, *'Broken Scale Invariance in the Standard Model'*, hep-th/0403039.

[2] C. Pilot and S. Rajpoot, *'Gauge and Gravitational Interactions with Local Scale Invariance'*, in *Proc. 7th Mexican School of Particles and Fields and 1st. Latin American Symposium on High-Energy Physics* (VII-EMPC and I-SILAFAE - Dedicated to Memory of Juan Jose Giambiagi), Merida, Yucatan, Mexico, 1996, ed. J.C. D'Oliva, M. Klein-Kreisler, H. Mendez (AIP Conference proceedings: 400, 1997), p. 578.

[3] H. Weyl, S.-B. Preuss. Akad. Wiss. **465** (1918); Math. Z. **2** (1918) 384; Ann. Phys. **59** (1919) 101; *Raum, Zeit, Materie, vierte erweiterte Auflage*: Julius Springer (1921).

[4] W.-F. Chang, J.N. Ng and J.M.S. Wu, *'A Very Narrow Shadow Extra Z-Boson at Colliders'*, [2]7406095005, hep-ph/0608068.

[5] W-F. Chang, J.N. Ng and J.M.S. Wu, *'Shadow Higgs from a Scale-Invariant Hidden $U(1)_s$ Model'*, hep-ph/0701254.

[6] K.A. Meissner and H. Nicolai, *'Conformal Symmetry and Standard Model'*, hep-th/0612165.

[7] H. Nishino and S. Rajpoot, *'Standard Model and SU(5) GUT with Local Scale Invariance and the Weylon'*, CSULB-PA-06-4, to appear in *CICHEP-II Conference Proceedings*, 2006, published by AIP.